\documentclass[12pt]{article}
\title{Foundations of Probability and Physics-3: Preface of Proceedings and Round Table on Quantum Foundations and Computing}

\author{International Center for Mathematical Modeling\\
in Physics, Engineering and Cognitive science\\
MSI, V\"axj\"o University, S-35195, Sweden}

\begin{document}

\maketitle

\begin{abstract}This preprint contains a detailed Preface to Proceedinngs of the International Conference  ``Foundations of Probability and Physics-3'' held in V\"axj\"o, Sweden, 7-12 June 2004; table of contents and round table. The main theme of the round table was {\it ``Fundamental problems
in quantum mechanics, probabilistic description of reality, and
quantum information.''} The topics that were specifically
discussed were that of Quantum Cryptography, Quantum computing,
and Quantum Macroscopic Structures. For each of these topics, the
participants were asked to discuss which are the crucial Quantum
features required among the following : violation of Bell's
inequality, Entanglement, Complementarity, and Interference of
Probabilities. Finally, the connection between Mental states and
Quantum states was discussed.
\end{abstract}

\bigskip

{\bf PREFACE}

\bigskip

This volume constitutes the proceedings of the International Conference  ``Foundations of Probability
and Physics-3'' held in V\"axj\"o, Sweden, 7-12 June 2004
(see also: http://www.msi.vxu.se/icmm/fpp3/).The organizing committee of the Conference included:
S. Gudder (University of Denver), H. Rauch (Atominstitut, Vienne), A. Yu. Khrennikov (V\"axj\"o University,
Sweden). The conference was supported by the Swedish Research Council, Profile Mathematical Modeling
of V\"axj\"o University, and the EU-network``Quantum Probability and Applications in Biology,
Economy, and Physics.'' This conference is one of a series of V\"axj\"o conferences on the foundations
of quantum mechanics (with emphasis on probability theory).\footnote{See: Proceedings of International
Conferences (ed.: A. Yu. Khrennikov): {\it Foundations of Probability and Physics,} Ser.  Quantum Prob.
White Noise Analysis, {\bf 13}, 201- 218(WSP, Singapore, 2001); {\it Quantum Theory: Reconsideration
of Foundations, }  Ser. Math. Modeling,v.2 (V\"axj\"o Univ. Press, 2002); {\it Foundations of Probability
and Physics}-2, Ser. Math. Modeling,v. 5 (V\"axj\"o Univ. Press, 2003); {\it Quantum Theory:
Reconsideration of Foundations}-2,  Ser. Math. Modeling, v. 10, V\"axj\"o Univ. Press, 2004; see
also http://www.msi.vxu.se/forskn/quantum.pdf, http://www.msi.vxu.se/forskn/mattebok5.pdf} The
conference started with an  opening lecture by Alain Aspect, who presented the conventional
viewpoint on the EPR-Bohm experiment, Bell's inequality and experimental tests of Bell'sinequality.
\footnote{There was an interesting deviation from the usual presentation: Instead of simultaneous
 measurements of commuting observables on parts of a composite quantum system, he considered
 conditional measurements: i.e., given the first subsystem as a condition, what state does the
 second subsystem exhibit. However, he did not regard this as a new viewpoint on the relation between
 realistic and quantum models. He was sure that this is just a convenient way to present the EPR-Bohm
 experiment.} 
 
 This first talk induced stormy debates, which where stopped only by the announcement that
 participants were about to miss lunch. Debates continued during lunch. There were many questions and
 strong comments from people who doubt the conventional interpretation of Bell's inequality. In particular,
  A. Khrennikov  pointed out that it seems that the problem which is well known in experimental
   circles under the rubric of  ``detector efficiency,'' is in fact, essentially a more general problem
   of ``matching'' measurements taken on two parts of composite systems. In particular, time-factors play
   a crucial role. It seems, that it is a fundamental problem, which will not be solved in a purely experimental framework. A. Aspect disagreed with such a viewpoint and expressed the belief that the problem of detector efficiency will besolved in the future via technological advancements.

   There was a heavy discussion between A. Aspect and I. Volovich. The latter pointed out that the conventional description of the EPR-Bohm experiment (in Bell's formulation) does not contain space as a variable at all. I. Volovich claimed that without additional experiments on the role of spacial separation of the detectors in EPR-Bohm experiments,Bell's conclusions are not justified. A. Aspect responded to Volovich's contention by pointing out that suitable experiments have been done; and, they did not demonstrate substantial influence by such separation on the  observed correlations.

   During this lunch discussion involving A. Aspect, A.Khrennikov, I. Volovich and A. Grib,  A. Khrennikov asked about arguments supporting the anti-photon interpretation of, say, Willis E. Lamb and Alfred Lande,  who, although among the founders of quantum mechanics, nevertheless totally rejected Einstein's photon. A. Aspect presented a detailed recounting of attempts to use a classical description of light in quantum theory to explain experiments, which, he said, show that such attempts can not be successful. 
   
Andrey Grib recounted the origin of the Copenhagen interpretation of the wave functions (i.e., assigning a wave function to an individual quantum system, e.g., to a single electron, vice an ensemble). He pointed out that one of the strongest supporters of this interpretation was Vladimir A. Fock, and that even though Bohr himself had doubts about its consistency, he, Fock, demonstrated to Bohr inconsistency in the Einsteinian ensemble interpretation.

A. Aspect and A.Khrennikov discussed the so-called V\"axj\"o (contextual statistical realistic) interpretation of quantum mechanics. A. Khrennikov claimed that by taking into account dependence of probabilities oncomplexes of physical conditions (contexts) one can reconstruct the probabilistic structure of quantum mechanics. The latter was considered as just a special (complex Hilbert space) projection of a realistic pre-quantum model. A. Aspect supported the contextual viewpoint to quantum mechanics, but he expressed his doubts regarding the possibility to reconstruct quantum mechanics solely on the basis of classical (even contextual) probability theory. A. Aspect was certain that  negative probabilities would always arise.
   
A. Aspect's opening presentation provided considerable stimulation; and, it was discussed in numerous coffee-breaks and during the round table. It was the first in a series of excellent talks on experiments with quantum  systems. In particular, Oliver Benson presented results of single photon experiments  with applications to quantum information. His talk induced a discussion regarding the justification for the claim, that single photon experiments really have been done. H. Rauch spoke about the foundations ofquantum mechanics in light of quantum interferometry. One of his messages was, that in real experimentsit is actually impossible to split microscopic systems from macroscopic experimental devices, e.g., aneutron from the crystal in the interferometer. H. Takayanagi described a wide domain of experimental research at the NTT Corporation (Kanagawa, Japan). This talk, among others, emphasized how truly amazingthe breadth of advanced experiments in so many directions on quantum information theory is in fact. G. Jaeger described experimental research in quantum information at the Boston University Photonics Center concentrating on entanglement and symmetry in multi-qubit states.

   These experimental talks on quantum information matched very well with fundamental theoretical talks on foundations of quantum information, computing and cryptography. In particular, we can mention the lectures of Masanori Ohya and Noboru Watanabe on characterization of quantum entangled states, quantum entropy and capacity; Igor Volovich onthe role of space in quantum cryptography; Sebastian Ahnert on optical implementation of Hardy'sparadox; Jan-{\AA}ke Larsson on the possibility of a coincidence-time loophole. There was also a series of talks on probabilistic foundations of quantum mechanics and measurement theory, e.g., Stan Gudder ---fuzzy quantum probability theory, Andrei Khrennikov ---contextualprobability theory and contextual interpretation of interference of probabilities, Olga N\'an\'asiova---independence on orthomodular lattices, Paolo Rocchi ---reversibility and irreversibility of stochastic systems, Inge Helland ---quantum theory as a statistical theory under symmetry and complementarity, Marcos Perez-Suarez --- subjectivist Bayesian approach to quantum information.

   As always, much was said about Bell's inequality. Luigi Accardi and Massimo Regoli presented graphical results from numerical experiments employing a classical, local realistic model reproducing the EPR-Bohm correlation function. These graphical results demonstrated clearly the essence of their so-called chameleon effect. This effect induces losses of particles due to influence by measurement devices. Therefore, it seems that the old bet\footnote{In 2002 at Madeira, after a good lunch, we (L. Accardi, R.Gill, V. Belavkin, I. Helland and I) took a walk on the beach where L. Accardi and R. Gill made a bet onthe possibility of a computer simulation (local and realistic) reproducing the EPR-Bohm correlation function. Others (V. Belavkin, I. Helland and I) were chosen to serve as the jury. The first bet was for 1000 Euro. Next year the ante was raised to  2500 Euro. Polemics between Accardi and Gill, however, diverted the dispute to an ancillary issue: namely, the correctness of the computer program used by Accardi-Regoli.} between L. Accardi and R. Gill should be adjudicated in favor of Accardi, as it really is possible to reproduce the EPR-Bohm correlations. The chameleon effect is local, and chameleons arereal. The only problem is, that some chameleons die, and these chameleons would not be taken intoaccount. The latter possibility was not discussed at Madeira, but Accardi's model does fulfill reality and locality conditions. But R. Gill might, of course, reply that he had in mind experiments without dead chameleons and so on. 
   
   One may say, that this story really is just about detector efficiency. However, it would be unreasonable to reduce the chameleon effect to only a matter of detector efficiency, because it involves a more general problem. In fact, a similar issue was discussed in the talk by Guillaume Adenier and Andrei Khrennikov, who reproduced the EPR-Bohm correlation function by taking into account the influence ofpolarization beam splitters. Adenier also designed an experimental scheme to test for the effects of such an influence. I would like to comment on these investigations by saying, that in fact they match very well with Bohr's thesis, that the whole experimental arrangement must be taken into account. Moreover, I think that the influence of devices plays the fundamental role, and they will not be eliminated through technological developments. This viewpoint was defended in our paper with I. Volovich at the V\"axj\"o conference two years ago.\footnote{A. Yu. Khrennikov, I. V. Volovich, Local Realism,Contextualism  and Loopholes in Bell`s Experiments. Proc. Conf. {\it Foundations of Probability andPhysics}-2, Ser. Math. Modeling , vol. 5, 325-344, V\"axj\"o Univ. Press, 2002; quant-ph/0212127.} It is interesting, that in fact people (using the conventional interpretation of Bell's inequality) have very strong disagreements with N. Bohr, but still, many people criticizing Bell's arguments actually are just follow N. Bohr. In any case, I think that until we are able to present at least a computer simulation ofthe whole EPR-Bohm experiment, which would model effects of all measurement devices (source, polarization beam splitters, phase shifters and so on), it is too early to come to definite conclusions.

   In his talk, W. Philipp criticized existing derivations of Bell's inequality. He considered two main approaches to derive Bell-type inequalities: 1) measure-theoretical; 2) frequency. It was emphasized,that one could proceed in the first approach only by assuming (as J. Bell did) that there exists a joint probability distribution of observables. This assumption has no real physical justification, since some observables are incompatible. This argument was discussed many times during previous conferences by L.Accardi, W. De Baere, W. De. Muynck, A. Khrennikov. The new feature was the rediscovery by W. Philipp ofan old paper by the Russian probabilist Vorobjev, who described random parameters which do not have ajoint probability distribution; he used such parameters in game theory. Thus, there is not hinge specially `quantum' in such situations. W. Philipp also criticized frequency derivations, because, in his opinion, one should not mix, into the same arithmetic expression, statistical data from different (even incompatible) experiments. This thesis was strongly criticized by practically all participants(including experimentalists).

   In principle, I presented similar arguments a few years ago\footnote{Khrennikov A.Yu., {\it Interpretations of Probability.} VSP Int. Sc. Publishers,Utrecht/Tokyo, 1999 (second edition, 2004).} (such a viewpoint was introduced  by W. De Baere already in  the 70's). However, I have been disappointed by the fact, that I am unable to find examples in the macro world that exhibit effects similar to those seen in quantum mechanics and caused by such data mixing. 
   
   There were many interesting talks on philosophical problems in quantum mechanics, e.g., Arcady Plotnitsky on complementarity, quantum probability and information; Gerard Emch on models of dynamicsin mathematics and theoretical physics, Harald Atmanspacher on epistemic and ontic realities. Theo Nieuwenhuizen showed that, contrary to a rather common opinion, one can  construct realistic modelsfor many quantum measurements. He also criticized strongly the orthodox Copenhagen interpretation and gave various arguments in favor of the so-called statistical interpretation of quantum mechanics. In a discussion after his talk, Luigi Accardi and Andrei Khrennikov pointed out that, although in general they support his views, the situation is not so simple. In particular, A. Khrennikov emphasized that explaining interference of probabilities in the statistical approach is a complex matter.  Ingemar Bengtsson gave the talk on finite geometries, polytopes and double stochastic matrices. Luigi Accardipointed out the connection of these structures with some complex problems in quantum probability. Andrey Grib discussed the appearance of quantum non distributive structures in macroscopic games and possible applications of such games. There were a few talks on applications to acoustic and signal analysis (Thomas Biro, Borje Nilsson and Sven Nordebo) and to financial mathematics (Roger Pettersson).
   
   A strongly `contrarian' presentation was made by A. F. Kracklauer, who first surveyed several critical studies of Bell's analysis, including Edwin Jaynes' argument that Bell misapplied the chain rule for conditional probabilities, and an argument originated by Brody et al. in the early 1970's, to the effect that data satisfying the conditions of derivation of Bell inequalities, physically can not be acquired. He then proceeded to present candidate, local-realistic numerical models (unrelated to the chameleoneffect) that exhibit EPR correlations. Altogether, this threefold complementarity between experimental,theoretical, and philosophical talks was very fruitful and hopefully stimulated discussions further than the conference boundaries, until next year.
   
   \medskip
   
Andrei Yu. Khrennikov,
   
Director of International Center for Mathematical Modeling

In Physics, Engineering and Cognitive Sciences

\newpage

{\bf ROUND TABLE}

The main theme of the round table was {\it ``Fundamental problems
in quantum mechanics, probabilistic description of reality, and
quantum information.''} The topics that were specifically
discussed were that of Quantum Cryptography, Quantum computing,
and Quantum Macroscopic Structures. For each of these topics, the
participants were asked to discuss which are the crucial Quantum
features required among the following : violation of Bell's
inequality, Entanglement, Complementarity, and Interference of
Probabilities. Finally, the connection between Mental states and
Quantum states was discussed.

\section{Quantum Cryptography}

\bigskip

{\bf Gregg Jaeger:} The no-cloning theorem is the essence. If
perfect cloning were possible then eavesdropping would be possible
and the special physical security of quantum key distribution
would be lost. Bell's theorem  does not play that crucial a role,
but without Bell's theorem what becomes of Quantum Mechanics? In
any case, there can be a practical benefit in using entanglement
such as in the Ekert protocol.

{\bf Luigi Accardi}: Yes, but no-cloning theorem is just a
theorem. It's more Heisenberg uncertainty principle that plays a
crucial role. There is no first kind measurement for photons.

{\bf Andrei Khrennikov}: I agree that the no-cloning theorem is
just a theorem. Any theorem is valid only for a fixed system of
axioms. The no-cloning theorem was proved in the conventional
(Dirac-von Neumann) axiomatic of quantum mechanics. This axiomatic
provides a rather special interpretation some Hilbert space
structures. In particular, it is crucial that the {\it
superposition} of states can be considered as a superposition of
states for an individual system. During the last hundred years
numerous arguments have been presented both for and against the
conventional axiomatic. One cannot exclude however that another
axiomatic of quantum mechanics could be created (with another
interpretation of the superposition principle), and in which the
no-cloning theorem would no longer be valid.

{\bf Igor Volovich}: It is in any case possible to make an
approximate cloning. The main danger actually comes from an attack
from space.

{\bf Jan-Ake Larsson}: Bell inequalities and no-cloning theorem
are important. What is important is that the devices introduce
something quantum into the play. The violation of Bell
inequalities provides a test of the devices as being truly quantum
or not.

{\bf Andrei Grib}: In some versions of quantum cryptography breaking of Bell's
inequalities plays a crucial role. If there is  intervention of Eve into
communication of Alice and Bob they can learn it because of validity of Bell's
inequalities.If they are broken then Alice and Bob are sure in the secrecy of
communication.  Breaking of Bell's inequalities shows that no Eve read their
message.

{\bf Karl Svozil}: Complementarity is the key. Propositional
structures fitting complementarity. A type of quantum cryptography
with classical objects and a kind of complementarity. Non
distributive.

{\bf Arkady Plotnitsky}: Complementarity preserves all the basic structures
of quantum mechanics and hence it preserved the no-cloning theorem.

{\bf Oliver Benson}: Single photons are crucial (violation of Bell
inequalities is not). Quantum repeaters for long distances are
needed and therefore quantum entanglement or teleportation.

{\bf Luigi Accardi}: I think one should distinguish the case of
weak coherent state cryptography and the case of entangled state
cryptography: in the former case security is only statistical
(although with reasonably high probability); in the latter the
crucial role is played not by the impossibility of cloning, but by
the impossibility of interacting with the photon without breaking
the entanglement.

{\bf Andrei Khrennikov}:  If violation of Bell inequalities are
not that crucial, then why do all talks on quantum cryptography
start precisely with Bell inequalities?

{\bf Gregg Jaeger}: Quantum Mystique!

\section{Quantum computing}

\bigskip

{\bf Andrei Khrennikov}:  Suppose we can go beyond quantum
mechanics and create a realistic pre-quantum model, then what
becomes of quantum computers?

{\bf Al Kracklauer}: Quantum computing exploits superposition of
states. This is okay for ensembles, but what about for ontological
single systems in such a superposition?  Experiments always
involve ensembles. Is superposition really needed for quantum
computing?

{\bf Andrei Khrennikov}: If one uses the ensemble interpretation
of quantum mechanics and the corresponding interpretation of
superpositions, then it is really impossible to explain the origin
of quantum parallelism in computations. If everything can be
described by purely classical probability distributions then the
advantages of quantum algorithms over classical vanishes. In this
case the whole quantum computing business would only remain as a
model demonstrating the possibilities of coherent manipulation of
quantum states.

{\bf Arkady Plotnitsky}: Quantum Computing indeed depends on superposition.

{\bf Hans Grelland}: Molecules and chemistry depends exactly on
this idea of cat states. The geometrical form of orbitals, which
are wave functions and not just probability densities, determine
the chemical properties of the molecule Thus the wave function is
demonstrated to be real by the chemical behavior of individual
molecules.

{\bf Al Kracklauer}: Quantum tomography yields just a
reconstruction. If one reconstructs a Beethoven symphony from
signals of an incomplete set of filters, the result would be far
from the original, and might well have nonintuitive properties.

{\bf Karl Svozil}: Quantum parallelism is just a
metaphor---Young's double slit experiment can be considered to
have this feature as well. The problem is that there is no
powerful enough algorithms. The existing ones are ingenious but
not extremely convincing. There are many important NP-complete
problems, and we have only at our disposal two algorithms which
are not in this class.

{\bf Stanley Gudder}: I agree. Superposition and entanglement are
crucial. Quantum cryptography is today technology, quantum
computing is but for the future.

{\bf Gregg Jaeger}:  Grover's algorithm does not need
entanglement, however. Of course decoherence comes into play for
all algorithms.

\section{Quantum Macroscopic Structures}

{\bf Andrei Khrennikov}:  What is the upper bound for cat states?
Anthony Leggett for instance sees no limit; Anton Zeilinger did
already beautiful experiments showing interference with heavy
atoms.

{\bf Igor Volovich}: Quantum Macroscopic Structures exists
already: superfluids, superconductors, etc. There is no problem in
creating a quantum macroscopic structure, and quantum mechanics is
actually the only way to explain these phenomenon.

{\bf Andrei Khrennikov}: But if there are no interferences here,
how come quantum mechanics be the only possible theory for these
phenomenon?

{\bf Gregg Jaeger}:  GRW put forward one. But where does this
collapse occur? Nobody has been able to answer that.

{\bf Helmut Rauch}: There are no boarders. Quantum objects are
combined with macro- objects (the measurement device). It just
becomes more and more difficult to shift the experimental boarder
to more macroscopic objects. It's a technological problem.

{\bf Andrei Grib}: Some people like Penrose and others have hope on quantum gravity
to solve the problem of the collapse. For any macroscopic system in spite of the
fact that it consists of quantum particles, movement of it's center of mass
must be described as movement of large mass, much larger than a Planckian
mass. If one will use quantum mechanics for this degree of freedom one can think
about quantum gravity. Some experiments were proposed in CERN by Ellis and
Nanoupulos to look for breaking of unitarity in these cases, but there are still
no definite results. Superstring theorists also have hopes on these experiments because
even before quantum gravity scale superstrings can manifest their existence due
to breaking of unitarity for macroscopic degrees of freedom.

{\bf Luigi Accardi}: This dualism Micro-Macro is not very
accurate. We should go further than historical tradition. There
are more than just two scales, a continuum in fact. Micro, Macro
(and Meso, which begins to be quite fashionable nowadays) are but
schematizations. The natural scales are infinitely many in a
double sense: (i) that there are infinitely many scales (of
magnitude - for masses, of distances, of energy, of wave length,
of time, ...); (ii) within each scale there are infinitely many
gradations. Any finite categorization necessarily introduces some
arbitrariness, in the sense that the boundary between two scales
is not sharply defined. In the case of the micro-macro scale this
was already pointed out by the Greek philosophers with the famous
``argumentum acervi''.

{\bf Karl Svozil}: But there are even some deterministic rules and
finite automata which could mimic complementarity.

{\bf Arkady Plotnitsky}: It depends on one's definition of complementarity.  Complementarity in
Bohr's sense applies to the measuring devices involved and  not to the
quantum objects ("particles") themselves.  As such, it establishes a
mutually exclusive character of certain measurements and predictions, for
example, in the EPR case, where what kind of measurement we perform on the
first object of the EPR pair strictly defines what we can or cannot predict
concerning the outcomes of possible measurements on the second.  If it is a
momentum measurement (manifest in the apparatus), we can only predict the
outcome of a momentum measurement, but never a position measurement, on the
second object (a measurement, again, to be manifest in the apparatus used to
verify the prediction), and vice versa.  How do you yourself define
complementarity?

{\bf Karl Svozil}: Complementarity can be characterized by the non
distributivity of the associated propositional structure; e.g., of
the quantum logic.

{\bf Andrei Khrennikov}: I think that the only quantum feature
that can be tested experimentally is the interference of
probabilities. In my contextual probabilistic approach,
interference is a consequence of a combination of statistical data
obtained for different complexes of physical conditions --
contexts ---, and therefore I totally agree with Arcady
Plotnitsky. In principle, combinations of contexts producing
interference of probabilities can appear in the macro-world. Such
combinations are described by quantum formalism. Therefore we
might find quantum effects in classical statistical mechanics,
economics, cognitive sciences, or even sociology.

\section{Is there any connection between
mental and quantum?}

\bigskip

{\bf Luigi Accardi}: To say that there is connection is a truism
once we accept quantum mechanics as our fundamental physical
theory. The hard problem is to deduce something nontrivial from
this statement. At the moment we can only say that it is a
beautiful dream and, in any case it seems to fascinate several
people! Also dreams play a role in the development of science ...

{\bf Oliver Benson}:  Biophysics developed enough to look for
smaller systems where there would be hints of such effects (and
don't start with such a complex system as a brain).

{\bf  Gregg Jaeger}:  The desired quantum effects in microtubules
have never been witnessed in the brain's environment,
nevertheless.

{\bf Helmut Rauch}:  There is a connection, yes. All our existence
is based on  quantum world. Quantum is not specifically needed to
understand this (mean field description is enough). However, I do
not see any field where quantum mechanics does not play a role.
Quantum mechanics is everywhere, stability of atoms, molecules,
biological cells, evolution of the universe, etc.

{\bf Andrei Khrennikov}:  Many years ago, Quantum Theory was
supposed to provide a new understanding of Mental. Now people in
cognitive sciences, neurophysiology, psychology are however very
upset. They spent ten years learning Quantum Theory for nothing,
and they are now very angry. I think that the root of the problem
is that attempts to connect Quantum and Mental are done in a very
direct way, by trying to reduce mental behavior to the properties
of quantum systems composing the brain (Penrose, Homeroff,...).

{\bf Olga Nanasiova}: Mathematical structures as orthomodular
lattices = Quantum (and quantum logic is important for this).

{\bf Luigi Accardi}: We should sadly acknowledge that, at the
moment, even classical physics and mathematics, applied to
concrete biological problems has been of a very limited
importance: I don't know of any mayor breakthrough in biology, not
to speak of medicine or of neurophysiology, where either
fundamental physics or sophisticated mathematics has played a role
comparable to the role now played in physics. For this reason i
believe that we should be very prudent in doing grandiose
statements on the possible role of quantum mechanics in the
explanation of the brain mechanisms. There is a concrete risk that
science will loose credibility if one promises much more than one
can effectively deliver. Let us not forget that the path from
first principles to final comprehension of individual complex
phenomena is very slow: it has historical times. For example a
first principle explanation of the mechanism of formation of
crystals is still lacking notwithstanding the efforts of several
first class physicists and mathematicians!

{\bf Marcos Peres-Suarez}: On the one hand, the question as proposed seemed to me to be
reductionistic and, on the other hand, that it was surprising that no
one had mentioned Pauli's collaboration with Jung and his fascinating
involvement with the notion of archetype, with the latter's role in
physics, and with the consideration of psychophysical parallelism. In
particular, there are some features in quantum theory which are
significantly reminiscent of the interplay between physics and the
psyche in the terms proposed by Pauli.

{\bf Andrei Grib}: Brain functions as a whole. There is some coherence in
the work of its different parts typical for a wave.The psychophysical problem as
the problem of interaction of consciousness and the physical substance of the
brain have much similarities with the role of measurement and collapse of the
wave function in quantum mechanics. One can speak about property like
complementarity in different states of consciousness especially in hypnosis,
dreams etc.However, it does not necessarily means that this quantum mechanics
of brain similarly to superconductivity occurs due to quantum physics of
neurons or some molecules of brain.It can arise because for such complex system
as brain  chance is not described by the standard probability theory but one
must use the formalism of the probability amplitude being more general
description of chance.

{\bf Andrei Khrennikov}: In collaboration with the group of
psychologists from the University of Bary we have performed
experiments on students (``incompatible tests'') and collected
data which showed the presence of interference of probabilities. I
interpret the results of these experiments as a preliminary
evidence of the possibility to use quantum formalism in mental
sciences. I do not think that it would be possible to derive
Mental from Quantum with a reductionist approach, but I am
convinced that Mental could and should be described by quantum
mathematics.

\newpage

{\bf CONTENTS}

\bigskip

L. Accardi,  Some loopholes to save quantum nonlocality\\

G. Adenier, A. Yu. Khrennikov, The quantum stalemate\\

A. E. Allahverdyan, R. Balian, T. M. Nieuwenhuizen,  The quantum
measurement process in an exactly solvable model\\

T. Aichele, U. Herzog, M. Scholz, O. Benson,  Single-photon
generation and simultaneous observation of wave and particle properties\\

V. A. Andreev, V. I. Man'ko,  Quantum tomography and verification of
qeneralized Bell-CHSH inequalities\\

H. Atmanspacher, H. Primas,  Epistemic and ontic quantum realities\\

I. Bengtsson,  MUBs, polytopes, and finite geometries\\

T. Biro,  On the vector Helmholtz equation in toroidal waveguides\\

G. G. Emch,  Not what models are, but what models do\\

S. Filipp, K. Svozil,  Tracing the bounds on Bell-type inequalities\\

L. Gregory,  Quantum filtering theory and the filtering
interpretation\\

H. H. Grelland,  A non-intuitionist's approach to the interpretation
problem of quantum mechanics\\

A. A. Grib, A. Yu. Khrennikov, G. N. Parfionov, K. A. Starkov,
Distributivity breaking and macroscopic quantum games\\

S. Gudder, Fuzzy quantum probability theory\\

T. Heinonen,  Fuzzy position and momentum observables.\\

I. S. Helland,  Quantum theory as a statistical theory under
symmetry.\\

K. Hess, W. Philipp,  Bell's theorem: critique of proofs with and
without inequalities\\

A. K. Higa, R. Pettersson,  Projection scheme for a reflected
stochastic heat equation with additive noise.\\

A. Hilbert,  Degenerate diffusions with regular Hamiltonians\\

G. Jaeger,  Entanglement and symmetry in multiple-qubit states: a
geometrical approach\\

A. Yu. Khrennikov,  Reconstruction of quantum theory on the basis of the
formula of total probability\\

A. Yu. Khrennikov, Y. I. Volovich,  Energy levels of hydrogen atom in
discrete time dynamics\\

A. F. Kracklauer,  Bell's inequalities and EPR-Bell experiments: are
they disjoint?\\

J.-A. Larsson, R. D. Gill,  Bell's inequality and the
coincidence-time loophole\\

H. J. Leydolt,  De Broglie's wavefunction and wave-particle dualism\\

V. Maksimov,  Skew algebras, existence and application\\

R. A. Mould,  Quantum brains: the oRules\\

I. D. Nevvazhay,  The relation between micro- and macro-worlds and
the problem of observation\\

B. Nilsson,  Inverse electromagnetic scattering for spiral grain in
trees\\

S. Nordebo, M. Gustafsson,  Multichannel broadband Fano theory with
applications in array signal processing\\

M. Ohya, T. Matsuoka,  Quantum entangled state and its
characterization\\

M. Perez-Suarez, D. J. Santos,  Bayesian intersubjectivity and
quantum theory\\

S. Perovic,  Recent revival of Schr\"odinger's ideas on interpreting
quantum mechanics, and the relevance of their early experimental critique\\

H. Rauch,  Quantum phenomena tested by neutron interferometry\\

P. Rocchi, G. Sh. Tsitsiashvili,  About the reversibility and
irreversibility of stochastic systems\\

K. Svozil,  On counterfactuals and contextuality\\

Y. O. Tan, B. L. Lan,  A stochastic mechanics and its connection
with quantum mechanics\\

N. Watanabe,  On general quantum mutual entropy and capacity\\

R. Woesler,  Problems of quantum theory may be solved by an
emulation theory of quantum physics\\

R. Woesler,  Quantum teleportation in spacetime, and dependent
clones with given probability\\

\end{document}